# Evaluation of 14 to 15-Year-Old Students' Understanding and Attitude towards Learning Einsteinian Physics

Tejinder Kaur[1], David Blair[1], Ron Burman[1], Warren Stannard[1], David Treagust[2], Grady Venville[1], Marjan Zadnik[1], Warwick Mathews[3] and Dana Perks[3]

[1]*University of Western Australia, Crawley, WA, Australia, 6009.*
[2]*Curtin University, Bentley, WA, Australia, 6102.*
[3]*Shenton College, WA, Australia, 6008.*

There is an increasing recognition regarding of the importance of introducing modern Einsteinian concepts early in science education. This study investigates the efficacy of an innovative educational programme "Einstein-First", which focuses on teaching Einsteinian physics at an earlier age than usual through the incorporation of appropriate hands-on activities. This paper presents an analysis of 14 to 15-year-old students' conceptualisation of Einsteinian physics and their attitudes towards science as a result of this programme. We have investigated the students' understanding of modern physics concepts after a term of 20 lessons. We report on two such 20-lesson programmes, one delivered in 2013 and a second, improved programme delivered in 2014; each to 50-60 students across two classes designated by the participating high school as "academically talented" students. We found, as expected, that the students' possessed little prior knowledge of Einsteinian physics. The significant improvement in the students' knowledge, as tested before and after the course, showed that they could comprehend Einsteinian physics at the level it was given. The findings also showed that the short programme improved students' attitude towards physics. While the male students initially showed greater interest in physics compared to their female counterparts, the female students showed a significantly increased interest in physics after the programme. There was no evidence found to indicate that the material presented was too complex or that the students were too young to grasp the topics discussed. Also, students' memory retention of Einsteinian physics concepts was tested in two different years, one class was tested after one year of the programme and the other was tested after three years of the programme. The results shows that the Einstein-First programme had a lasting impact on the students involved in the study.

## I. INTRODUCTION

Our current understanding of the universe is based on two theories of physics; the theory of gravity, also known as general relativity, and the theory of particle interactions or quantum mechanics. Because of Einstein's central role in both theories, we use the term "Einsteinian Physics" to distinguish these subject matters from classical Newtonian Physics. Einstein's key ideas include the relativity of time and the curvature of spacetime, as well as stimulated emission and Brownian motion [1, 2, 3]. Many others, including Bohr, Planck and Heisenberg contributed to this paradigm shift in scientific thinking – from Newtonian to so-called "modern physics".

Einsteinian physics is of immense importance in modern technology. A communications system, lasers, transistors, semiconductors, nuclear power and many more of today's technologies are based on Einsteinian physics. However, in many countries, including Australia, the high school science curriculum is focused mainly on Newtonian physics with Einsteinian concepts generally introduced only as special advanced topics [4]. It is often believed that Einsteinian physics is too difficult to introduce to this age-group.

Introducing Einsteinian physics into the school science curriculum is one of the challenges for science education [5, 6, 7]. There are difficulties in teaching and learning Einsteinian physics in schools such as mathematical complexities, a transition from classical physics to relativistic physics [8, 9, 10]. Hence these topics are usually left later for university education and thus most students are not exposed to these ideas. Researchers also found that students struggle to understand these concepts at the university level because of their existing knowledge of Newtonian physics [11]. This could be one reason for students dropping out of physics courses. Thus, it is possible to assume that an earlier introduction of Einsteinian physics concepts may help students to better understand the reality of the universe and prepare them to acquire more details on these topics in their university studies.

Attitude towards science is one of the major concerns in science education which could be influenced by many factors such as: perception of the science teacher, teaching style, teaching environment, motivation towards science, enjoyment of science and attitudes of friends or classmates [12, 13]. An effective teaching environment and teaching method have a positive impact on students' attitude towards science [14]. Previous research showed that experiments using simple materials helped students to reflect on events that occur in nature and activities enhanced their cognitive skills and hence had positive effects on students' attitude [15, 16]. Researchers also found that worksheets [17] and science activities carried out with simple and inexpensive materials [18, 19] improved the students' attitudes towards science. The Einstein-First project considered these approaches while delivering lessons in the classrooms. Einsteinian physics concepts were taught through models and analogies by dividing students into groups.

In Australia, students' attitude towards science and the number of students studying science are decreasing [20]. Previous research showed that the students' attitude towards science decreased with increasing year level [21, 22, 23, 24]. The motivation for this project is early exposure to Einsteinian physics concepts will enhance students' attitude and interest in science. Many researchers are also considering the effects of curriculum on science attitudes [25, 26].

Globally, researchers are puzzling over factors which influence students' attitude towards science. Many of them concluded that gender has a significant effect [27, 28, 29], with studies reporting that males generally exhibit a more positive attitude towards science than females [20, 30]. The investigation of males and females' attitudes towards studying science has been popular work for many researchers for the past 30–40 years. The most common ways to examine males and females' attitudes towards science are by comparing their scores [31], their performance and ability [32] or interest in science [33].

Another study by Reid on males and females' attitudes towards science and physics in particular, showed that at the end of primary education, both genders were found to have positive attitudes towards science but towards the end of secondary education, a significant decline in females' attitude was observed [34]. Jones et al. also reported in their study, that high school appeared to be a critical time for science-related experiences. Therefore, it is important to focus on this period due to the more substantial differences in attitude between the two genders compared to their attitudes in middle school [35]. Reid also observed that females' interests were more inclined towards subjects related to social or daily life whereas their male counterparts showed their interest towards mechanical relevance. He also mentioned that it is important to create a more balanced physics syllabus by considering both male and female interests. Lorenzo et al. and Pollock et al. found that active pedagogies or in-class interactions might help to reduce the gender gap in attitude towards science [36, 37].

To evaluate learning, students' conceptual understanding of science is always regarded as one of the most important research issues [38]. Chang et al. mentioned in their study that "tests have been regarded as not only an important means for assessment, but also as an influential method for developing students' conceptual understanding" [39]. Another research showed that tests are powerful tools in assessing students' conceptual understanding as well as improving their learning [40]. Previous research showed that the use of models and analogies in the learning and teaching process helps students to engage themselves in learning and developing conceptual understanding of a particular topic [41, 42]. According to Posner et al., for conceptual change in students' understanding, a new conception should be understandable, believable and provides new possibilities or ideas [43]. Treagust et al. suggest that students need to use their prior knowledge and ideas to make judgements in new conceptions [44].

Researchers from a number of countries are working on the feasibility of introducing Einsteinian physics concepts into the school science curriculum through various approaches [45, 46, 47, 48]. The most common ones are "thought experiments", simulations [49], physical models [50] and analogies. Previous research shows that many students have the ability to understand some qualitative concepts of Einsteinian physics. In Australia, for example, research has shown a significant improvement in students' understanding and interest in Einsteinian ideas. The students did not consider that their age was a barrier in learning modern physics [51]. Haddad and Pella reported that Year 6 students could understand the concepts of relativity at the knowledge level [52]. Johansson and Milstead said that the introduction of Uncertainty principle helps students to understand the mysterious nature of quantum mechanics [53].

Our research programme, named Einstein-First, has been developing and testing approaches to the teaching and learning of Einsteinian physics across the school curriculum. We introduce modern concepts such as curved space, gravity, a geometry of curved space, the universality of free fall, dual nature of light, quantum interference and diffraction, the uncertainty principle, the photoelectric effect, etc. To make these topics accessible to school students, we have developed an extensive curriculum based on the use of physical models and analogies. These models support activity-based learning and enable a systematic development of the fundamental concepts of quantum mechanics and special and general relativity. The curriculum material is not presented here but is described in papers [54, 55].

Our work is based on the contention that;

a) Students deserve to be taught our current best understanding of the nature of the universe in which we live.
b) It is possible to learn the modern paradigm at a young age using appropriate and carefully designed materials.

c) Student attitudes to physics will improve if they learn modern concepts before being introduced to the useful tools of Newtonian physics.

## A. Purpose of the study

The purpose of this study is to assess how two Year 9 classes in 2013 and 2014 respond to a 20 lesson programme designed to introduce the concepts of Einsteinian physics.

## B. Research objectives

In particular, the objectives of this study are to discern the following:

1. What are the students' pre-instructional understandings of Einsteinian physics?
2. How do the students' understandings of Einsteinian physics change as a result of the Einstein-First teaching and learning programme?
3. How do the students' attitude and interest in physics change as a result of the Einstein-First teaching and learning programme?
4. How much did these high school students retain from what they had learned in Einsteinian programme?

The paper is structured as follows: Section II of this paper decribed the methodology including the description of questionnaires used in this study, student sample, structure of the programme, data analysis, validity and reliability of the questions. Section III is based on the research results. In this section, we first presented the students' results from conceptual pre/post-tests followed by the attitudinal pre/post-results. At the end of this section, students retention results were discussed.

## II. METHODS
## A. The study

The Einstein-First teaching and learning programme that is the focus of this research consists of twenty lessons with activities, worksheets and questionnaires. In this study, four questionnaires were developed by the researcher. These questionnaires developed through a process described in subsection E below.

The questionnaires are as follows:
1. *Conceptual pre-questionnaire*
   This questionnaire, consisting of nine short and multiple-choice questions, was given at the start of the programme; it was designed to assess the students' prior knowledge of Einsteinian concepts (see Table II). The questionnaire focused on the students' understanding of curved spacetime, geometry on curved space, gravity, light and the uncertainty principle. The marking criteria are described below.
   - 0 - No response, incorrect or unsure;
   - 1 - Correct short answer or a correct multiple-choice response;
   - 1 - Correct Yes/No answer; or
   - 2 - Correct Yes/No with an explanation.

2. *Conceptual post-questionnaire*
   This questionnaire was given at the end of the programme, and was designed to assess any conceptual change after having participated in the programme. The pre- and post-questionnaires had identical questions and were marked using the above criteria.

3. *Attitudinal pre-questionnaire*
   Attitudinal pre-questionnaire consisted of nine questions (see Table III). This questionnaire was designed to assess the students' general attitude towards physics. All the answers were given based on a Likert scale (strongly disagree to strongly agree). The Likert scale marks were quantified according to a 1 – 5 scale, normally 1 for strongly disagree and 5 for strongly agree but reversed where necessary.

4. *Attitudinal post-questionnaire*
   This questionnaire had identical questions to the pre-questionnaire and was designed to observe any change in the students' attitude towards modern physics. This questionnaire was marked using the scale given above.

5. *Views on Einsteinian programme post-questionnaire*
   A set of five questions requiring short responses was developed to determine the students' views on Einsteinian physics (see Table IV). This questionnaire was only given at the end of the programme.

6. *Delayed retention test*
   A set of 9 questions was designed to assess the students' retention about Einsteinian concepts (see Table V). This was administered after 1 year with one class and after 3 years with another.

Each questionnaire was distributed at the start of a standard period before and after the programme, and students were given 20 minutes for completion. The same questionnaires were used in both the 2013 and 2014 studies.

## B. The structure of the programme

Twenty lessons were structured with carefully designed models and analogies to assist with students' understanding. Generally the lessons were structured according to the following format:
1. First 15 minutes of each lesson were dedicated to introducing and presenting material for the lesson,
2. The next 15 minutes to group activity,
3. And the last 15 minutes on a class discussion and the completion of worksheets.

The lesson plans were designed with the intention of making Einsteinian physics interesting and engaging for students of varying academic abilities. The activities and materials that supported this programme are described in references 21 and 22. They included lessons using woks and magnetic posts for studying curved space geometry, a lycra sheet space-time simulator for studying gravity and curved space effects, toy "photons" based on Nerf guns for studying the particle nature of light, water ballons for studying the concept of terminal velocity for developing the concept of the speed limit of the universe and also the universality of free fall, laser experiments for diffraction and quantum interference, and videos of single photon interference.

**C. Student sample**

The sample of students used in this research consisted of 120 Year 9 students from Shenton College in Perth, Western Australia. Students were from an academically talented stream, but most had not encountered Einsteinian concepts prior to the study. The programme was run in two years; 2013 and 2014. The final sample of this study consisted of 102 students due to incomplete questionnaires, which reduced to about half for the retention tests undertaken in 2014 and 2017.

**D. Data analysis procedures**

Data was processed using Excel to obtain means, standard deviations etc. A paired samples t-test was used to evaluate any difference in the students' conceptual understanding after the programme. To analyse attitudinal pre/post questionnaires, we combined data for positive responses according to the Likert scale given above.

**E. Validity**

Validity is defined as the degree to which test scores precisely measure the proposed idea. The validity of the conceptual and attitudinal questionnaires were investigated under the following three questions:
1. Do the questions encompass every topic we wish to teach the students and have these topics been addressed in the literature?
2. Are the students able to interpret the questions as they are meant?
3. Do educational experts agree that the questions are appropriate?

An extensive review process described below was used to ensure validity.

**1.** *Content and Literature validation*

In designing the conceptual and attitudinal questionnaire, we took the approach that the assessment of students' learning and attitude should be based on the topics we covered in the specially-designed programme on Einsteinian physics concepts. Only content-related questions were asked. As discussed in Section I there has been very little research on the teaching of the fundamental concepts of Einsteinian physics. Hence it is not surprising that some of the conceptual questions have not been previously reported. We only found 3 out of 9 conceptual questions in existing literature [18] as many of them were designed according to the need of the programme. We chose 3 out of 9 attitudinal questions from the literature, meaning they are validated questions [56].

*2. Student interpretation validation*

The language of the questions should be simple and clear so that the students would interpret questions as asked. Some of the questions such as "Does space have a shape?" and "What is light?" were naturally ambiguous but the ambiguity was explained to the students so that they understood for example that light was not the opposite of heavy and that space refered to the space between your hands or the space of the classroom, rather than outer space. Most other questions used were directly related to topics covered in the programme.

*3. Expert validation*

Draft questions were reviewed by the authors who include experienced physicists and educators. Each question was discussed and refined. The conceptual questions were reviewed in relation to a large data base of physics questions. All questions were redrafted and reviewed a second time before finalising them for use in this study.

**F. Reliability**

For the student attitude questionnaire we used internal consistency as a measure of reliability. In order to investigate the internal consistency, we used Cronbach's alpha [57]. We calculated the value of Cronbach's alpha as 0.89, which indicates that all the questions are highly reliable.

The following section presents the results obtained from the questionnaires.

**III. RESULTS**

We shall now discuss the findings of the two 20-lesson programmes. The first programme was introduced in 2013. In 2014 the lesson plans and activities were refined. The skills of the presenter Tejinder Kaur had significantly improved in 2014 through the experience of the first year. There was one additional difference between the programmes in 2013 and 2014. In 2014 students undertook a mid-programme test which may have helped to reinforce their learning.

We first present an overview of results for conceptual learning. We divided the test scores into three bands: low achieving ( 0 – 40%), mid range (40 – 80%) and high achieving (81 – 100%) as shown in Table I. These are presented separately for each year of the programme. Results highlight the high level of learning achieved, especially in the second year. The individual questions are presented in section C where we analyse results for each question.

TABLE I. Students' statistical results from 2013 and 2014.

| Distribution of scores | Percentage of students (2013) | | Percantage of students (2014) | |
| --- | --- | --- | --- | --- |
| | Pre | Post | Pre | Post |
| Low range (0 – 40%) | 76% | 0% | 91% | 2% |
| Mid range (41 – 80%) | 24% | 80% | 9% | 10% |
| High range (81 – 100%) | 0% | 20% | 0% | 88% |
| Mean | 27% | 73% | 23% | 91% |
| SD | 17.5% | 12.6% | 14.6% | 10.3% |
| Paired t-test | $t(44) = 20.1$ | | $t(56) = 30.3$ | |
| Results | Statistically significant | | Statistically significant | |

In the second year 88% of the students were in the high achieving band. It is also interesting to note that the standard deviation (SD) between the pre and post-tests reduces despite the increased mean scores. This implies that there was rather uniform uptake of the concepts taught. In section A we discuss details of the 2013 result, followed by the 2014 results in section B. In an attempt to understand some of the higher scores obtained in knowledge pre-test, we asked the classes where they had learnt Einsteinian concepts. Most indicated that they had learnt from the internet or TV.

### A. Students' overall results in conceptual learning from 2013

Figure 1 presents the results of 45 students who completed their pre and post questionnaires conducted in 2013. Results are presented in the ascending order of percentage in the pre-test. In the pre-test, 93% of students scored less than 50%. Just 11 students scored more than 40% in the pre-test. This may reflect some prior knowledge as discussed above. However we noted that none of these students gave requested explanations against the five yes/no answers in the questionnaire.

The highest score achieved in the pre-test was 65% by student number 45. After the programme, this student achieved a 100% score, as did student number 30. In the post-test, only one student (student number 11) scored less than 50%. These results demonstrate that the students significantly improved their level of knowledge in Einsteinian physics after undergoing the programme.

The 13 students who scored less than 10% in the pre-test increased their mean score to 68%. The 11 students who scored more than 40% in the pre-test achieved a mean final score of 81%. While the top achieving scores in the post-test are weakly correlated with the scores in the pre-test, overall the class score is relatively uniform. No student showed a decreased score indicating that every student improved their knowledge of Einsteinian concepts. The significance of the improvement is clearly statistically significant.

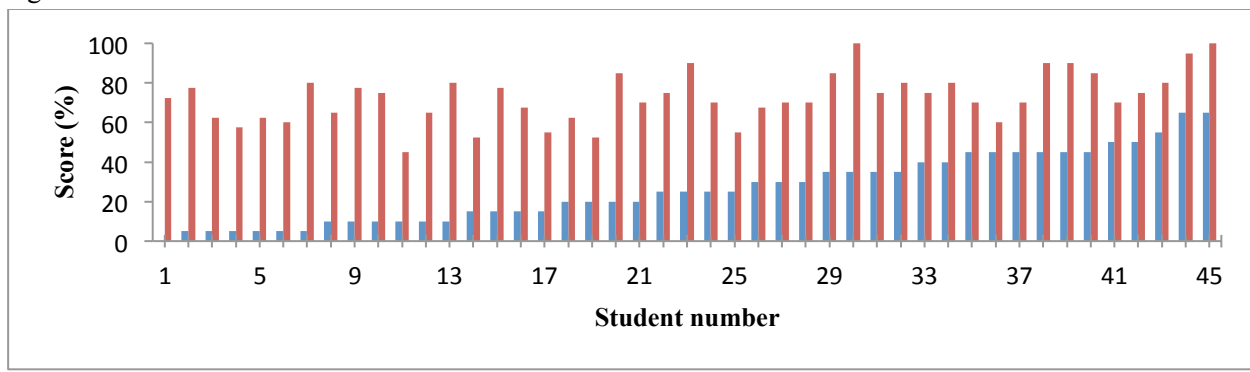

FIG. 1. Conceptual understanding test results for 2013. The histogram, ranked in order of the pre-test score, shows that almost all students achieved a high level of conceptual understanding, independent of their pre-programme understanding which was generally low.

### B. Students' overall results in conceptual learning from 2014

The results from the pre/post questionnaires in 2014 are presented in Figure 2, following the format of Figure 1. First it is interesting to note that the distribution of initial scores is similar to that of the class of 2013. However it is also clear from this data that the post-test results are significant higher.

In the pre-test, 8 students who scored less than 10% increased their mean score to 94%. This exceeds the mean score of the class, and is consistent with the evidence from 2013, indicating that the learning outcomes are rather independent of student prior knowledge. Interestingly the two students (student number 1 and 2) who had the lowest scores in the pre-test achieved 100% in the post-test. The 5 students who scored more than 40% in the pre-test achieved a mean final score of 94%. In comparison in 2013, 11 students in this category who achieved 81% in the post-test. In the post-test, only one student scored below 50% (student 24). The paired sample t-test confirmed that there was a statistically significant increase in scores from pre-test to post-test as shown in Table I.

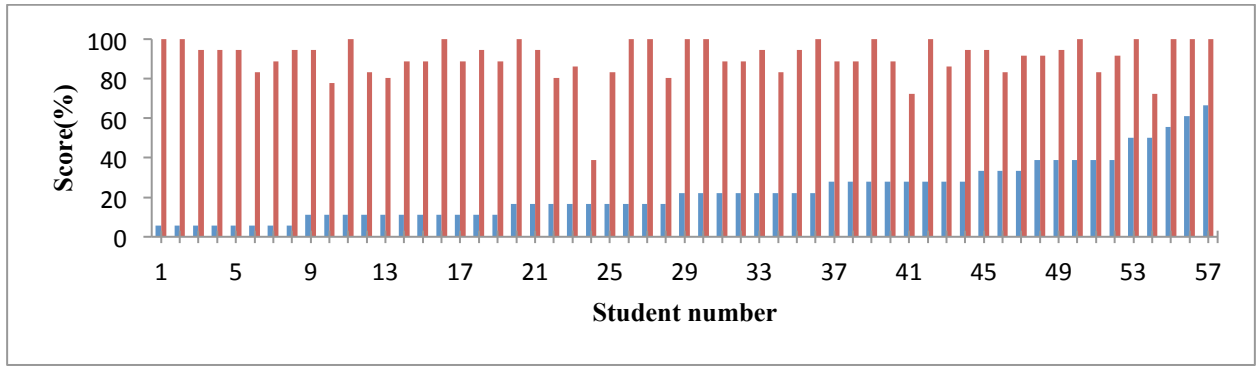

FIG. 2. Conceptual understanding test results for 2014. The data in Figure 1 clearly shows similar behaviour to that obtained in the previous year, but notably the post-test results are significant higher.

## C. Analysis according to each question

In Table II, we show the scores obtained from conceptual pre-test and post-test for the entire cohort of students over both years, for each of the nine conceptual knowledge questions. The general improvement in student knowledge as already summarised in Table I is apparent, but there is also significant scatter. Two questions (CQ 5 and CQ 7) stand out with higher pre-test marks. We will discuss the reasons for this in sections 2 and 3. Also two questions (CQ 3 and 9) have very low marks. Analysis of individual answers showed that for the first question, students understood light as something that allows you to see or removes darkness, without having any sense of its nature. Students had no pre-knowledge that light is either an electromagnetic wave or a stream of photons. The second question relates to a much more sophisticated concept – the quamtum uncertainty principle – for which the low score is not unexpected.

TABLE II. Students' result for each question before and after the programme. The results show that students improved their conceptual learning in every concept after the programme. The maximum improvement was seen in the concepts of experimental geometry CQ 1 and CQ 2 for which the post-test score was 100%.

| Conceptual Question (CQ) | Questions asked in the pre/post-test | Pre-test (%) | Post-test (%) |
|---|---|---|---|
| 1. | Can parallel lines meet? | 15 | 100 |
| 2. | Can the sum of the angles in a triangle be different from 180 degrees? | 17 | 100 |
| 3. | What do you mean by the term "Light"? | 3 | 92 |
| 4. | Does space have a shape? Circle Yes or No. How could you measure the shape of space? | 34 | 90 |
| 5. | If you weighed an object on a supersensitive balance, would the balance register a different weight if you heated the object up? | 47 | 82 |
| 6. | How could you tell if a ruler is straight? | 26 | 73 |
| 7. | In the absence of air resistance, (like in a huge vacuum tank or on the moon) if we drop a hammer and a feather, which one of them will touch the ground first? | 66 | 100 |
| 8. | List the names of at least four types of electromagnetic radiation. | 30 | 81 |
| 9. | A person claims on Facebook that he has made a perfect microscope that is so accurate that the exact position of an atom can be measured. Could this claim be plausible? | 4 | 80 |

*1. Questions on curved space geometry: CQ 1, CQ 2, CQ 4 and CQ 6*
Questions CQ's 1, 2, 4 and 6 were designed for testing students learning about curved space geometry. This topic was taught through a range of activities:
   a) Trajectories of toy cars: To see that parallel lines can meet, students studied trajectories on a curved elastic membrane.
   b) Experimental geometry on woks: Students had learnt experimental geometry by drawing triangles of different sizes on woks (cooking utensil) with magnetic posts. They found that the sum of the angles of a triangle is not equal to $180^0$ on curved areas.
   c) Straight lines on a curved surface: Students had learnt how we could define straightness using laser light or a tightly stretched string on a curved two-dimensional surface.
The mean scores of CQ 1, CQ 2, CQ 4 and CQ 6 were 100%, 100%, 90% and 73% respectively. Clearly, students achieved excellent understanding of curved space geometry.

*2. Question on mass-energy equivalence: CQ 5*
Question CQ 5 was chosen to test student learning about the equivalence of mass and energy; $E = mc^2$. There were two parts to this question, yes or no, and an explanation. In the pre-test almost all students answered yes, but failed to give any explanation. We are surprised by this result because there is little public awareness of this subtle effect on mass caused by the presence of the thermal energy. In the lessons, this concept was taught. Specifically students had been asked to calculate the increased mass of a phone battery when it is charged. The 82% mean in the post-test indicated significant understanding of this concept.

*3. Question on free fall: CQ 7*
Question CQ 7 on free fall is a topic that is quite widely covered in space media, for example a beautiful BBC video by Brian Cox of free fall in a NASA vacuum tank and video from the moon landing. Thus the high pre-test score is not surprising. Having undertaken free fall experiments during the programme, students scored 100% in the post-test.

*4. Questions on light and electromagnetic waves: CQ 3 and CQ 8*
Questions CQ 3 and CQ 8 were designed to test student understanding of the nature of light and electromagnetic waves. As already discussed, in the pre-test students demonstrated minimal knowledge of the physical nature of light, the CQ 8 score shows knowledge of electromagnetic waves. Following the programme, they had learnt that light is an electromagnetic wave with a dual nature (wave/particle duality). To study the wave nature of light, students undertook simple interference experiments. To understand photons, we used small plastic projectiles to mimic photons so that they could study an analogy of the photoelectric effect [22]. After the programme, 92% of the students were able to correctly describe the nature of light (CQ 3) compared to an initial 3%.

*5. Question on the uncertainty principle: CQ 9*
Question CQ 9 was intended to test understanding of uncertainty principle which makes a perfect microscope impossible. Students had learnt about this using "Nerf gun photography" in which they had observed the disturbance of a balloon which impcted by Nerf gun photons. They had learnt that photons disturb the objects they measure. The 80% post-test score shows that most students had grasped this concept.

## D. Attitude questionnaire

A questionnaire consisting of nine questions was given to the students to evaluate their attitudes toward physics. The questions used are listed in Table III. The questionnaire employed the Likert scale and the students were asked to rate identical questions before and after the sessions. The results from the two years (i.e. 2013 and 2014) were combined to give a total questionnaire population of 102 students comprising 57 males and 45 females. For this analysis, the positive answers ('Agree' and 'Strongly Agree') were combined and expressed as a percentage of the population. We examined two factors in analysing the answers. The first was to determine the general attitude of the students towards physics. Secondly, we examined any changes in the students' responses before and after undergoing the programme. We analysed the answers relating to the questions by dividing them into four categories, which are explained in the following paragraphs.

TABLE III. A set of attitudinal questions asked in pre and post-tests in 2013 and 2014.

| Attitude Question (AQ) | Attitudinal questionnaires asked in pre and post-test |
|---|---|
| 1. | I think physics is an interesting subject. |
| 2. | I prefer to learn physics through hands-on activities. |
| 3. | I enjoy learning new concepts and ideas. |
| 4. | I enjoy trying things out at home and/or telling my family about school science activities. |
| 5. | I think hands-on activities help me understand and remember new ideas much better than if they are just from books and formal lessons. |
| 6. | The things that Einstein discovered are important for modern technology. |
| 7. | I like doing mathematical calculations. |
| 8. | Understanding scientific ideas are more important than just memorising facts. |
| 9. | I enjoy science excursions and would like to have more of them. |

*1. Students' attitude toward activity-based learning and calculations*

To investigate students' attitude toward activity-based learning, AQ 2 (I prefer to learn physics through activities), AQ 5 (I think doing activities helps me understand and remember new ideas much better than if it is just from books and lessons) and AQ 9 (I enjoy science excursions and would like to have more of them) were combined. Students' attitude towards doing calculations was assessed through AQ 7 (I like doing calculations). In the analysis of AQ 2, AQ 5 and AQ 9 we combined students 'agree' and 'strongly agree' responses together. Amongst the group of students surveyed there was a strong preference for activity-based learning as shown in Figure 3. There was an overwhelming agreement from both male and female students that they prefer to learn through activities and excursions and that this is not only enjoyable, but they learn more by doing so. However, it is clear from Figure 3 that the male students were more inclined towards learning physics through activities and excursions. It is also interesting to note that the male students were more enthusiastic to participate in activities in the classroom or outside the classroom and more willing to help in setting up experiments compared to their female counterparts.

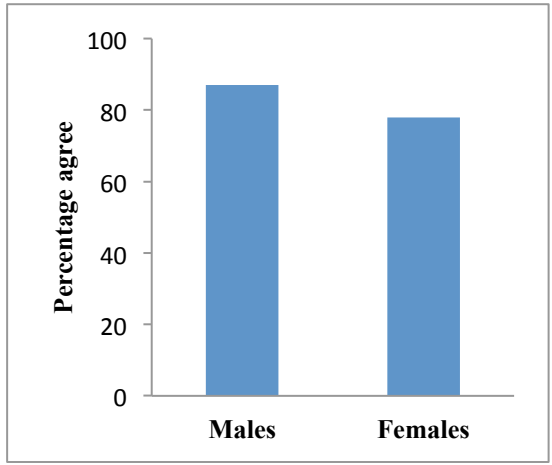

FIG. 3. Students' attitude to learning physics through activities and excursions.

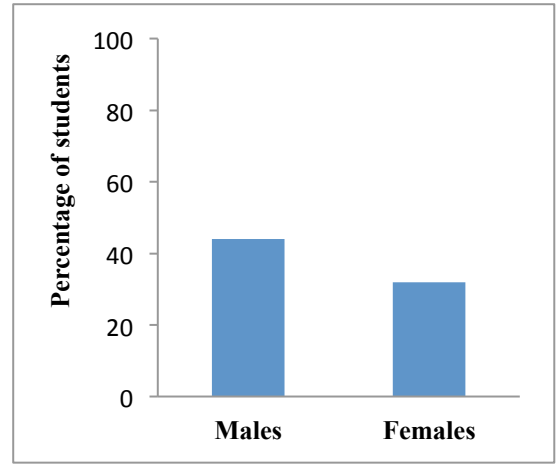

FIG. 4. Percentage of students who enjoy doing calculations.

Figure 4 shows students' attitude towards doing mathematical calculations. Most students, and significantly more females than males, do not enjoy calculations and prefer not to be taught calculations-based physics. Only 44% of males and 32% of females showed interest in doing calculations.

### 2. Students' interest in physics

Student interest in physics was evaluated by AQ 1 (I think physics is an interesting subject) and AQ 4 (I enjoy trying things out at home and/or telling my family about school science activities). The results from these questions are shown in Figure 5 and Figure 6.

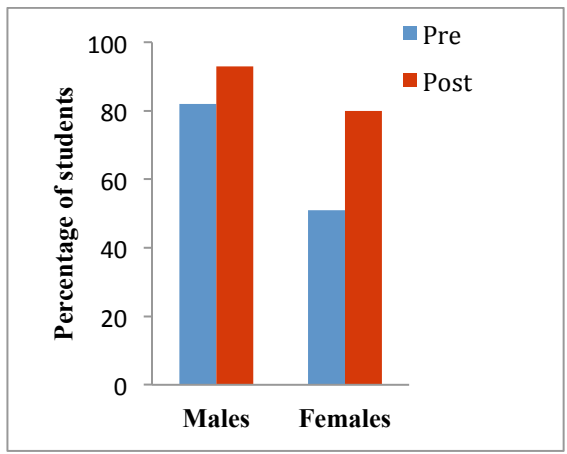

FIG. 5. Percentage of students who think physics is an interesting subject.

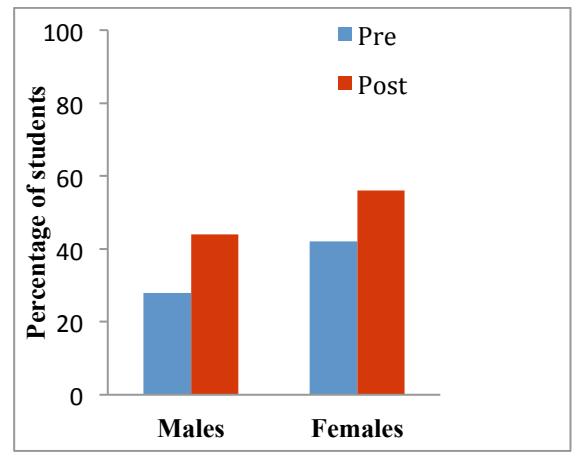

FIG. 6. Percentage of students who enjoy trying things out at home and/or telling family about school science activities.

Figure 5 indicates that initially 80% of males and 50% of females find physics interesting. After the Einstein-First sessions, interest amongst females increased to be almost equal to the initial interest of the males. The degree of interest is explored further in AQ 4 which tried to measure how much their interest in physics translates beyond the classroom. The positive response rate as shown in Figure 6 is significantly less than that in AQ 1 indicating that maybe the degree of interest is not always high enough to take outside the classroom. However, there is a significant increase in the interest acknowledged by both males and females after the sessions. The results indicate that the students' interest in physics improved significantly after the Einstein-first sessions.

*3. Relevance of physics*

To measure student awareness of the significance of Einsteinian physics in the modern world, we created AQ 6 (The things that Einstein discovered are important for modern technology). There was overwhelming agreement with this question with a slight increase evident after the sessions. The result is shown in Figure 7. This tells us that the students were already aware of the importance of Einsteinian physics before they started the programme. This may indicate that they recognised that the works of Einstein are relevant in understanding modern technologies so as might be expected, there was hardly any change in the students' answers after the programme.

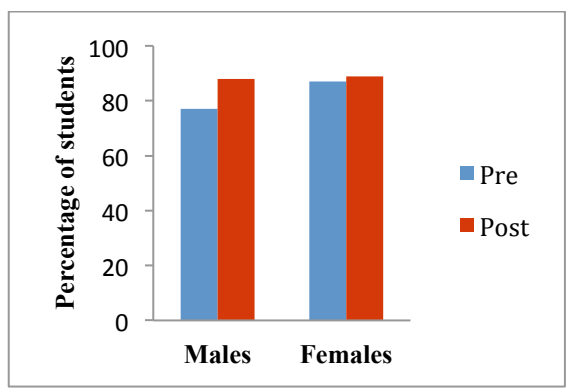

FIG. 7. Most of the students were aware of the importance of Einsteinian physics. This marginally increased through the programme.

*4. Learning concepts*

Questions AQ 3 (I enjoy learning new concepts and ideas) and AQ 8 (Understanding scientific ideas are more important than memorising facts) evaluated the students' attitudes toward the learning of concepts and ideas rather than information and factual material. The results of these two questions were combined and shown in Figure 8.

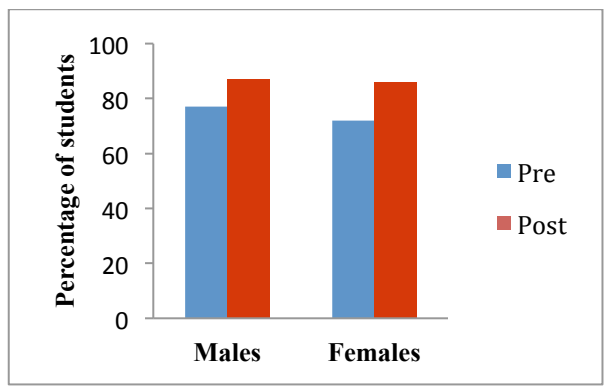

FIG. 8. Both boys and girls enjoyed new ideas and concepts.

Figure 8 shows that both males and females already had a positive attitude to learning new concepts. There was an increment observed after the programme. Students prefer to be exposed to concepts and ideas rather than facts and information. There was a significant increase in students' positive response after the sessions.

**E. Students views about Einsteinian physics**

We designed a questionnaire to evaluate students' opinion about the Einstein-First programme. The questions which were assessed are presented in Table IV.

TABLE IV. A set of questions asked to the students in the post-test only.

| Question Number (EQ) | Questions asked in the post-test to know students views about Einsteinian physics | Percentage who agree |
|---|---|---|
| 1. | The hands-on activities were very helpful to you for understanding the concepts. | 88% |
| 2. | I like doing calculations on $E = mc^2$. | 57% |
| 3. | I like doing calculations on gravitational lensing. | 34% |
| 4. | I would like to learn more about Einsteinian physics. | 70% |
| 5. | I think Einsteinian physics should be included in the curriculum. | 63% |

The analysis revealed that 88% of students responded positively to the use of hands-on activities (EQ 1). They agreed that hands-on activities helped to clarify Einsteinian concepts.

There were two lessons in the programme which were based on mathematical calculations. One lesson related to $E = mc^2$ and the second lesson was based on the more mathematically complex gravitational lensing calculations. In response to the EQ 2, 'I like doing calculations on $E = mc^2$', we observed that 57% agreed and 20% disagreed. For the question EQ 3, 'I like doing calculations on gravitational lensing', 34% students agreed and 28% disagreed. Students found calculations based on energy-mass equivalence easier when compared to gravitational lensing. Most of the students were aware of this famous equation before the programme but did not understand it.

To evaluate the students' eagerness to learn more modern physics, we used EQ 4, 'I would like to learn more about Einsteinian physics' 70% of students agreed with this question, while only 7% disagreed. In response to the EQ 5, 'I think Einsteinian physics should be included in the curriculum', 63% of students agreed and only 5% of students disagreed while others were unsure.

**F. Delayed retention test**

We created another questionnaire to examine whether the Einstein-First programme had a lasting impact on the students involved in the study. The students' retention of Einsteinian physics concepts was tested in two different years. The class attending this programme in 2013 was tested after one year. The 2014 class was tested after three years. We asked 9 questions covering all the concepts taught in the Einsteinian physics programme. The questions are given in Table V below.

TABLE V. Questions used to test retention of students after Einsteinian programme. This test was conducted twice, one in 2014 for students who attended this programme in 2013 and the other in 2017 for Year 9 students who attended this programme in 2014.

| Retention Question (RQ) | Questions asked |
| --- | --- |
| 1. | List three key concepts you learnt in the Einsteinian physics programme. |
| 2. | Have any concepts you have learnt in physics conflicted with what you learnt in the Einsteinian physics programme? Circle yes or no. If yes, please explain your answer. |
| 3. | The Einsteinian physics programme increased my interest in physics. Please circle your answer.<br>a) Strongly disagree<br>b) Disagree<br>c) Neutral<br>d) Agree<br>e) Strongly agree |
| 4. | Describe how space is related to gravity. |
| 5. | List two properties of space that contradict Euclidean geometry. |
| 6. | Describe an activity or experiment from Einsteinian physics programme that showed light behaving as particles (photons)? |
| 7. | Describe an activity or experiment from Einsteinian physics programme that showed how light can act as a wave? |
| 8. | In physics, scientists can estimate the mass of a galaxy by looking at the light from something far away behind it. What basic idea underpins this? |
| 9. | A typical kettle has a power of about 2000 Watts. If you fill it with water and switch it on for one minute, how much will the total mass of the kettle + water have increased?<br>[Hint: Energy = power x time; $E = mc^2$] |

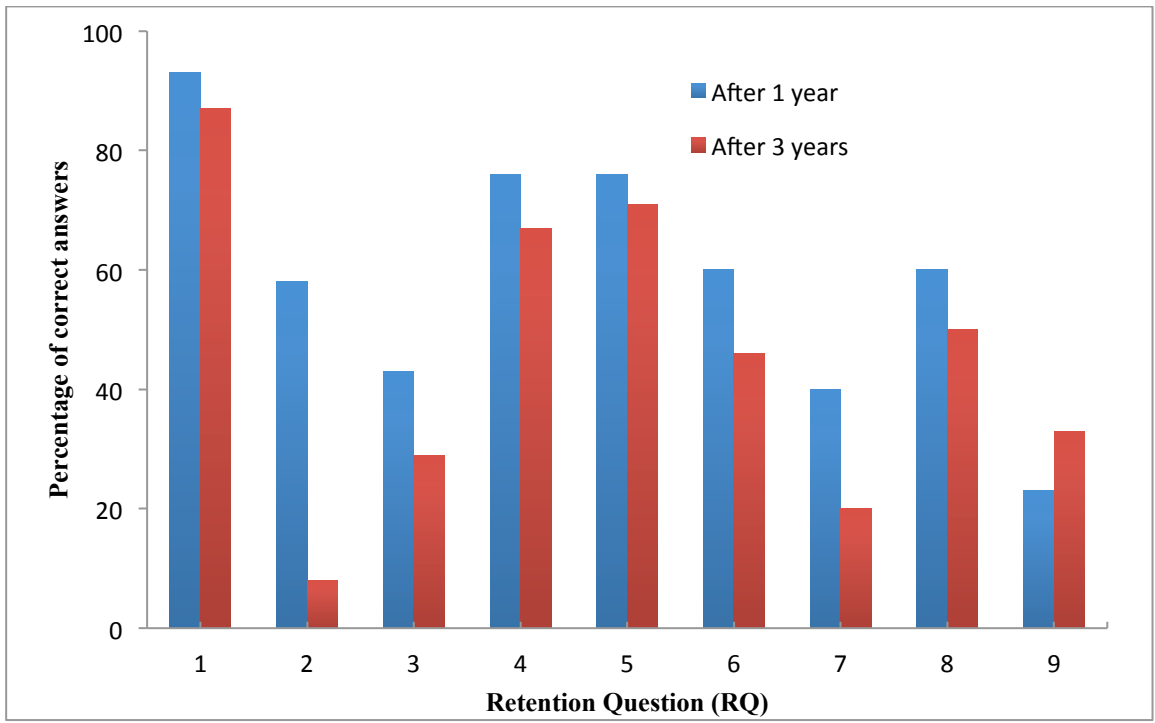

FIG. 9. Students' positive results for delayed retention test after one and three years of the Einsteinian physics programme. The results are displayed according to each question asked in the retention test.

*1. RQ 1: 2014 and 2017 results*
As shown in the figure, most of the students were able to recall three key concepts (RQ 1) they had learnt in the Einsteinian physics programme. Their answers included "space is curved", "geometry of curved space" and "light comes as photons". We noticed that most of the students who did well in the questions related to non-Euclidean geometry in the post-test, still remembered those concepts. Many students mentioned the concepts they were taught using Nerf gun such as the particle nature of light.

*2. RQ 2: 2014 and 2017 results*
When we asked about any contradictions between the concepts they had learnt in the Einsteinian programme and in Year 10 physics, 58% of students mentioned that they found some conflicts between the Einsteinian physics programme and the the concepts learnt in their current Year 10, while only 8% of Year 12 students found some contradictions after the programme. The reason for Year 12 students may be they have learnt some Einsteinian phsyics concepts in their Year 12 physics. The most common contradiction they mentioned is about the Newton's explanation of gravity.

*3. RQ 3: 2014 and 2017 results*
RQ 3 recorded the students' responses about their interest in physics. After one year of completing the programme, almost half of the students agreed that the Einsteinian physics programme had made them more interested in physics while only 29% of those who completed the programme 3 years ago gave a positive response.

*4. RQ 4 and RQ 5: 2014 and 2017 results*
After one year of the programme, 76% of students were able to explain how space is related to gravity (RQ 4) and 67% of students retained this concept after three years of the programme. RQ 5 tested their retention about non-Euclidean geometry. A year later, 76% of students remembered the properties of space that contradicts Euclidean geometry while 71% recollected it after three years.

*5. RQ 6 and RQ 7: 2014 and 2017 results*
A year after the programme, 60% of students could recall the name of the activity they did to understand the particle nature of light and this number decreased to 40% after 3 years of the programme. Furthermore, a year later, 40% of the students mentioned the name of the experiment they did to understand the concepts of quantum interference and diffraction whereas after 3 years of the programme, only 20% were able to recall the activity.

*6. RQ 8 and RQ 9: 2014 and 2017 results*
In 2014, 60% of students were able to recall the concept that scientists use to estimate the mass of a galaxy (RQ 8) and half of the students (50%) responded correctly in 2017 retention test. Our programme is based on conceptual understanding and simple mathematics. We only asked one mathematical calculation question (RQ 9) to the students. We found that a year after the programme, only 23% of students could solve the numerical problem. Conversely, 33% of the students tested 3 years later solved the same problem. We suggest that this improvement is due to their increased maturity and mathematical skills development.

## IV. Conclusion

The findings of our research into the Einstein-First enrichment programme shows that Year 9 students had a very positive response to the Einsteinian concepts. The idea of teaching modern physics with the help of appropriate activities appears to be very effective. Before the introduction of this programme, students had little knowledge of Einsteinian physics, while afterwards, a dramatic improvement in knowledge was observed. Students were able to comprehend concepts which are usually considered difficult, or beyond their capabilities. Of particular significance is the change in student attitudes to physics. This is a positive result in a traditionally male centred subject. While only a 30 - 40% of students reported sufficient interest in physics to bring it home, the girls scored higher than the boys, although the improvement factors were actually lower for the girls (boys 1.6 and girls 1.3). The retention results show that the Einsteinian programme had a long impact on students knowledge. Students were able to recall the activities they had learnt one and three years ago.

The results of this study show that we can achieve our aim if we teach students with simple and effective activities. This way, we can maintain students motivation until the end of their high school education. Eventually, this positive attitude may lead them to pursue a degree in physics.

We believe that the research provides substantial evidence that high school physics should be taught with a much greater emphasis on the Einsteinian physics which underpins modern technology and our best understanding of the universe.

## ACKNOWLEDGEMENT

This research was supported by a grant from the Australian Research Council, the Gravity Discovery Centre and the Graham Polly Farmer Foundation. The authors are grateful to the teachers Warwick Mathew, Dana Perks and Laura Ashbolt , relief teachers and students who participated in this study.